\documentstyle[prb,aps]{revtex}

\begin{document}

\draft

\title{The square-lattice spiral magnet Ba$_2$CuGe$_2$O$_7$ in an in-plane
magnetic field.}

\author{A. Zheludev, S. Maslov, and G. Shirane}

\address{Physics Department, Brookhaven National  Laboratory,
Upton, NY 11973-5000, USA.}

\author{Y. Sasago, N. Koide, and K. Uchinokura}

\address{Department of Applied Physics, The University of Tokyo,\\
 7-3-1 Hongo, Bunkyo-ku, Tokyo 113, Japan.}

\author{D. A. Tennant and S. E. Nagler}
\address{Oak Ridge National Laboratory,
Building 7692, MS 6393,  P.O. Box 2008, Oak Ridge, TN 37831, USA.}

\date{\today}

\maketitle

\begin{abstract}
The magnetic structure of Ba$_2$CuGe$_2$O$_7$   is investigated by
neutron diffraction  in magnetic fields applied along several directions
in the $(a,b)$ plane of the crystal. In relatively weak fields,
$H\lesssim 0.5$~T, the propagation vector of the spin-spiral rotates to
form a finite angle with the field direction. This angle depends on the
orientation of ${\bf H}$ itself.  The rotation of the propagation vector
is accompanied by a re-orientation of the plane of spin rotation in the
spiral. The observed behaviour is  well described by a continuous-limit
form of a free energy functional that includes exchange and
Dzyaloshinskii-Moriya interactions, as well as the Zeeman energy and  an
empirical anisotropy term.
\end{abstract}

\narrowtext
\twocolumn

\section{Introduction}
Incommensurate magnetic structures in insulators are as a rule caused by
a competition between two or more magnetic interactions. In the classic
description of a spiral spin structure in MnO$_2$ Yoshimori attributed
the  spiral spin arrangement to a competition between  two distinct
antiferromagnetic (AF) Heisenberg exchange interactions in the
crystal.\cite{Yoshimori59} We shall refer to this kind of spiral state
as ``exchange spiral''. An alternative mechanism is realized in a few
systems with non-centric crystal structures. Isotropic exchange terms in
the Hamiltonian compete with Dzyaloshinskii-Moriya (D-M)
interactions.\cite{Dzyaloshinskii57,Moriya60} The former are
proportional to $({\bf S}_{1} \cdot {\bf S}_{2})$, i.e., the scalar
product of interacting spins, and thus favor a collinear state. The D-M
term is given by a {\it vector} product and is usually written as $({\bf
D}\cdot [{\bf S}_{1}
\times {\bf S}_{2}])$, where ${\bf D}$ is the so-called
Dzyaloshinskii vector that characterizes the oriented bond between spins
${\bf S}_1$ and ${\bf S}_2$. D-M interactions favor a relative angle of
$90^{\circ}$ between spins. Unlike isotropic  exchange, D-M coupling is
of relativistic origin, i.e., is a result of  spin-orbit interactions,
and ties the spin system to the underlying crystal lattice via orbital
degrees of freedom.  The best known examples of real compounds where
this second scenario (``relativistic spiral'') is realized are the
MnSi\cite{Ishikawa76,Ishikawa77,Ishikawa-Shirane77,Ishikawa84} and
FeGe.\cite{Wilkinson76,Lebech89} Recently we have observed an
incommensurate spiral spin structure in Ba$_{2}$CuGe$_2$O$_7$  and
explained it within this
framework.\cite{Zheludev96-BACUGEO,ZM97-BACUGEO-L,ZM97-BACUGEO-B}

While the ground states in both models (``exchange '' and
``relativistic'') may be similar, under certain conditions they behave
very differently in applied magnetic fields. If only isotropic exchange
interactions are present, the direction of magnetic field is irrelevant
to the spin structure: the spin space may be freely rotated without a
change in interaction energy and the spins are not tied to any
particular direction in the crystal. For ``relativistic'' spiral magnets
the spin structure in an external field ${\bf H}$ in general may be
strongly dependent on the relative orientation of ${\bf H}$ and the
Dzyaloshinskii vectors, although in several special cases of
high-symmetry crystal  structures this is not so. In FeGe for example,
the interactions of each spin with its neighbors are characterized by
Dzyaloshinskii vectors pointing in all three equivalent orthogonal
directions. In this situation applying a very small magnetic field
always leads to a re-orientation of the spin rotation plane normal to
the field direction, and, consequently, the propagation vector becomes
aligned along the field. The length of the magnetic propagation vector
is to a good approximation field-independent.

In materials of lower crystal symmetry, as for example in hexagonal
CsCuCl$_{3}$\cite{Mino94,Schotte94,Ohyama95,Jacobs} and RbMnBr$_{3}$
\cite{Heller94,Zhitomirsky95}, or tetragonal Ba$_{2}$CuGe$_2$O$_7$, the
spin rotation plane can not be freely re-oriented without affecting the
Dzyaloshinskii energy. As a result, the magnetic structure undergoes
drastic changes for certain field geometries, and the length of the
propagation vector becomes strongly field-dependent.  An extreme example
of this behaviour is the commensurate-incommensurate transition that we
have recently observed and theoretically analyzed for
Ba$_{2}$CuGe$_2$O$_7$.\cite{ZM97-BACUGEO-L,ZM97-BACUGEO-B} As a magnetic
field is applied along the unique $c$ axis of the tetragonal structure,
the initially uniform sinusoidal spin-spiral is distorted. In a non-zero
field it may be viewed as a soliton lattice, a regular arrangement of
anti-phase domain wall boundaries separating regions of commensurate
antiferromagnetic spin-flop phase. As the field is increased, the period
of the structure, given by the distance between solitons, increases and
diverges at some critical field, $H_{c}
\approx 2.1$~T, above which a commensurate spin-flop state is observed.

In this work we address the field dependence of the magnetic structure
of Ba$_{2}$CuGe$_2$O$_7$ in a magnetic field ${\bf H}$ applied in the
$(a,b)$ tetragonal plane. We show that the square-lattice spin
arrangement in this geometry allows  an almost unhindered re-orientation
of the spin plane  with only a very small change in the period of the
structure, much like in cubic FeGe. The principal difference between
FeGe and Ba$_{2}$CuGe$_2$O$_7$ is that in the former material the
propagation vector rotates towards the  direction of applied field. We
find that  for Ba$_{2}$CuGe$_2$O$_7$  the propagation vector tends to
form a finite angle with ${\bf H}$, that depends on the orientation  of
${\bf H}$ itself. The rotation of the propagation vector with increasing
field is continuous. The observed behaviour is found to be in very good
agreement with theoretical predictions based on a 2-dimensional
generalization of the simple Ginzburg-Landau energy functional that we
have previously employed to describe the Dzyaloshinskii transition in
Ba$_{2}$CuGe$_2$O$_7$.

\section{Experimental procedures}
A single crystal of Ba$_{2}$CuGe$_2$O$_7$, the same sample that was used
in our previous work, was studied in a neutron diffraction experiment at
the High Flux Isotope Reactor at Oak Ridge National Laboratory, on the
HB-3 triple-axis spectrometer. The sample, roughly
$4\times4\times4$~mm$^{3}$ in volume, was mounted in a small aluminum
container where it was secured with compressed Al-foil. The container
was in turn mounted on a precision micro-goniometer, that was used to
align the $(a,b)$ crystallographic plane in the scattering plane of the
spectrometer, prior to setting up the sample in the cryomagnet.

The magnetic field was produced by using a split-coil {\it
horizontal-field} superconducting solenoid. The  construction of the
magnet provides two pairs of windows for the incident and outgoing
neutron beams. The larger pair of windows, each window being
$\pm30^{\circ}$ wide, is oriented perpendicular to the field direction.
An additional narrower pair of windows ($\pm15^{\circ}$) is positioned
around the direction of magnetic field. Within the cryomagnet the sample
could be manually rotated around the vertical axis {\it in situ}. We
estimate the uncertainty in the initial alignment of the sample relative
to the field direction to be of the order of $1^{\circ}$. After this
initial setting the relative rotation of the sample in the magnet was
done with much higher accuracy, $\approx 0.1^{\circ}$, as it could be
directly followed by monitoring the in-plane Bragg reflections.
Measurements were done for several values of the angle $\psi$  between
the magnetic field and the $[1 1 0]$ direction, namely, for $\psi=$45.0,
50.0, 52.9, 57.5, 60.2, 61.7, 61.9, 71.5 and 75.2$^{\circ}$, the field
always being applied in the $(a,b)$ plane of the crystal. Although the
magnet is capable of much higher field strengths, stray fields are a
serious problem, and all the measurements were  done in $H \le 2$~T. The
temperature of the sample was controlled by a standard He-flow cryostat,
allowing us to perform measurements in the temperature range
$1.8$--$10$~K. As we will discuss in detail below, the magnetic
structure changes in applied fields, but this effect shows a very strong
field-hysteresis. To avoid this complication all the measurements were
done using field-cooling. The sample was first warmed up to $T=6$~K,
well above the temperature of magnetic ordering
$T_{N}=3.2$~K,\cite{Zheludev96-BACUGEO} the desired strength and
direction of the magnetic field were set, and only then was the sample
brought down to base temperature, $T=1.8$~K, where most of the
measurements were performed.

At all times the measurements were done in 3-axis mode to reduce the
signal to background ratio. $60'-40'-40'-120'$ collimations were used
with pyrolitic graphite (PG) (002) reflections for monochromator and
analyzer, with $14.7$~meV neutron energy and a PG filter positioned in
front of the sample.

\section{Experimental results}
The crystal and magnetic structures of Ba$_{2}$CuGe$_2$O$_7$ are
discussed in detail in our previous publications on the
subject,\cite{Zheludev96-BACUGEO,ZM97-BACUGEO-L,ZM97-BACUGEO-B} and only
the most essential features are reviewed here. The magnetcic Cu$^{2+}$
ions form a square lattice in the $(a,b)$ tetragonal plane of the
crystal. The principal axes of this square lattice, hereafter referred
to as the $x$ and $y$ axis, are along the $[1 1 0]$ and $[1
\overline{1} 0]$ directions, respectively. To complete the coordinate
system we shall choose the $z$ axis to run along $[0 0 1]$. In the
ordered phase (below $T_{N}=3.2$~K) the spins lie in the $(1,-1,0)$
plane and the magnetic propagation vector is $(1+\zeta,\zeta,0)$, where
$\zeta=0.027$. The magnetic structure is a distortion of a N\'{e}el spin
arrangement: a translation along $(\case{1}{2},\case{1}{2},0)$ induces a
spin rotation by an angle $\alpha=\case{\pi}{\zeta}\approx8.6^{\circ}$
(relative to an exact antiparallel alignment) in the $(1,-1,0)$ plane.
Along the $[1 \overline{1} 0]$ direction nearest-neighbor spins are
perfectly antiparallel. Nearest-neighbor spins from adjacent Cu-planes
are aligned parallel to each other.

 Obviously, domains with propagation vectors $(1+\zeta, \pm
\zeta, 0)$ are equivalent and always equally represented in a
zero-field-cooled sample. This can be seen in Fig.~\ref{cont}(a) which
shows a contour plot of elastic intensities measured at $T=1.8$~K on a
mesh of points around the $(1,0,0)$ position in zero field. We clearly
see 4 magnetic reflections symmetrically grouped around the AF zone
center. The two pairs of peaks, $(1\pm\zeta,\pm\zeta,0)$ and
$(1\pm\zeta,\mp\zeta,0)$, correspond to the two magnetic domains. The
propagation vector ${\bf q}$, measured relative to the AF zone center,
is strictly along the $[1 1 0]$ (or $[1 \overline{1} 0]$) direction. The
weak feature around $(1,0,0)$ is an artifact that is
temperature-independent and was previously identified as a multiple
scattering peak.\cite{Zheludev96-BACUGEO} Cooling the sample in a very
weak magnetic field ($\approx 10$~mT) always produces a single-domain
structure.

When a magnetic field is applied along the $[1 0 0]$ direction the
diffraction pattern begins to change. The propagation vector ${\bf q}$
starts to rotate in reciprocal space towards the direction of applied
field [Fig.~\ref{cont}(b)], eventually becoming perfectly aligned with
it [Fig.~\ref{cont}(c)]. Note that all this happens in rather small
fields, less than $0.3$~T. This is to be compared with the field along
the $c$ axis, $H_{c}\approx 2.1$~T,  that is required to induce the CI
transition studied previously. Since $H_{c}\mu_{B}\approx |{\bf D}|$, we
can conclude that for a horizontally applied field the rotation of ${\bf
q}$ occurs in fields that are an order of magnitude smaller than the
Dzyaloshinskii energy.

In the following discussion it is convenient to introduce notations for
some angles in our experiment (Fig. \ref{notation}). We shall denote by
$\psi$ the angle between the magnetic field ${\bf H}$ and the $[1 1 0]$
direction in the crystal. $\phi$ will stand for the angle between $[1 1
0]$ and ${\bf q}$. Using $[1 1 0]$ as a reference is very convenient
since i) $\phi=0$ for $H=0$ and ii) the principal axes of the
square-lattice arrangement of the magnetic Cu sites in
Ba$_{2}$CuGe$_2$O$_7$ are along $[1 1 0]$ and $[1 \overline{1} 0]$,
respectively. In these terms, for $\psi=\pi/4$, as $H$ is increased,
$\phi$ continuously increases from $0$ to $\pi/4$.

For several directions of the magnetic field we have performed careful
measurements of $\phi(H)$. The results are summarized in Fig.
\ref{angles}. Measurements with $\psi<61^{\circ}$ were done on the
magnetic satellites around the $(1,0,0)$ AF zone-center, and those
around $(2,1,0)$ were used for higher values of $\psi$. This was
necessary to work around the geometrical constraints set by the cryostat
windows. As mentioned,  for $\psi=\pi/4$ the propagation vector starts
out at $\phi=0$ for $H=0$ and rotates all the way towards the field
direction ($\phi=\psi=\pi/4$), as in the case of FeGe. For arbitrary
field direction though, while $\phi$ levels off above $H\approx 0.5$~T,
it does so at a smaller value, i.e., before ${\bf q}$ reaches the
direction of applied field. If we plot $\phi$ measured at a relatively
high field, $H=2$~T, we find that the saturation value for $\phi$ is
always $\pi/2-\psi$ (Fig. \ref{linear}). Thus at high fields the
$(1,0,0)$ vector always bisects the angle formed by the magnetic field
${\bf H}$ and the propagation vector ${\bf q}$. Note that in all cases
we have $\psi \ge \pi/4$. If the magnetic field forms a smaller angle
with the $[1 1 0]$ direction, the magnetic domain with propagation
vector $(1+\zeta_{h}, \zeta_{k}, 0)$, $\zeta_{h}\zeta_{k}>0$ is
destroyed, and the diffraction intensity completely shifts over into the
$\zeta_{h}\zeta_{k}<0$ domain, returning us to the situation when $\psi$
is effectively greater than $\pi/4$. This fact has been verified
experimentally.

An important experimental result is that for all directions and of the
applied magnetic field  the length of the magnetic propagation vector
$|{\bf q}|$ is only weakly field-dependent. This can be seen in
Fig.~\ref{trajectory} that shows the trajectory traced by the
propagation vector in reciprocal space as the magnetic field is
increased. For $\psi=\pi/4$ this trajectory is an almost perfect
circular arc that starts out at $\phi=0$ and continues all the way to
$\phi=\pi/4$ [Fig.~\ref{trajectory}(a)]. For arbitrary  $\psi$ we see
that $|q|$ is indeed slightly field-dependent and the trajectory has a
characteristic S-shape. This, however, is a weak effect, barely
detectable with the precision of our experiment, limited by the
$Q$-resolution of the spectrometer.

Having discussed the field-dependence of the propagation vector, we now
turn to that of the spin orientation. Typically to solve the magnetic
structure one has to measure the intensities of several Bragg peak. In
the present experiment, however, we  did not have this luxury. The
severe geometrical constraints imposed by the narrow cryostat windows
eliminate simultaneous access to several magnetic peaks.  Many of these
could be observed by rotating the sample within the magnet, however, in
this  case the direction of the magnetic field relative to the sample is
changed. Only in the case of $\psi=\pi/4$, thanks to the presence of two
orthogonal pairs of windows, could we simultaneously observe more than
one set of magnetic satellites, namely those around $(1,0,0)$ and
$(0,1,0)$. The intensities of these reflections are plotted against $H$
 in Fig.~\ref{intensity} for $\psi=\pi/4$. These data are not
sufficient to independently determine the magnetic structure for each
value of applied field. We can however assume that for all cases, just
as for $H=0$, the spin arrangement is a flat magnetic spiral, with the
normal to the spin rotation plane confined in the $(a,b)$ plane of the
crystal. The assumption is quite reasonable if we consider that in our
experiments $H\le 2$~T and $\mu_{B}H \lesssim 0.1$~meV,  so the Zeeman
energy is at least an order of magnitude smaller than the exchange
energy ($4JS\approx 1$~meV). A slight conical distortion of the planar
spiral structure is of course  inevitable in finite magnetic fields, but
the tilt towards the direction of ${\bf H}$ will remain very small. If
we denote by ${\eta}$ the angle between $[1 1 0]$ and $\hat{{\bf n}}$,
the normal to the spin rotation plane  ($\eta=\pi/2$ for $H=0$), the
intensities of magnetic satellites around $(1,0,0)$ and $(0,1,0)$ are
determined by the spin polarization factors in the neutron diffraction
cross-section:
\begin{eqnarray}
 I_{(100)}\propto 1+ \cos^{2} (\eta-\pi/4); \nonumber \\
 I_{(010)}\propto 1+ \sin^{2} (\eta-\pi/4). \label{inten}
\end{eqnarray}
From Fig.~\ref{intensity} we see that at high fields $I_{(100)}$
increases by roughly a third of its original value, while $I_{(010)}$
decreases by roughly the same amount. This observation is consistent
with $\eta=\pi/4$ at high fields, as could be expected: a sufficiently
strong magnetic field will  always align the spin rotation plane
perpendicular to itself, independent of the field direction. For
arbitrary  $\psi$ we therefore expect $\eta=\psi$  in sufficiently high
fields. Experimentally, for all directions of magnetic field studied,
this rule was found to be consistent with the observed intensity
increase in the satellites around $(1,0,0)$ or $(2,1,0)$ that occurs
upon increasing the magnetic field from $H=0$ to $H=2$~T.

As will be discussed in detail in the next section, theory predicts that
for {\it arbitrary} values of $H$ one has
\begin{equation}
\phi=\pi/2-\eta. \label{rule}
\end{equation}
Since $\phi(H)$ is directly measured in our experiments, we can plot the
expected field dependencies of $I_{(100)}$ and $I_{(010)}$  using Eqs.\
(\ref{inten}) and (\ref{rule}). These are shown in solid lines in
Fig.~\ref{intensity} and are reasonably consistent with experimental
data.

We can now summarize all the experimental results in a  short list of
``rules'': i) As the magnetic field is increased, ${\bf q}$ rotates in
reciprocal space for small $H$, but slows down and eventually stops for
$H\gtrsim 0.5$~T; ii)  $|{\bf q}|$ is  only weakly field-dependent; iii)
at high fields the $[1 0 0]$ direction bisects the angle formed by the
vectors ${\bf H}$ and ${\bf q}$; iv) at high fields $\hat{{\bf n}}$ is
pointing along the direction of applied field; and v) it appears that
for arbitrary $H$ the $[1 0 0]$ direction always bisects the angle
formed by the vectors ${\bf q}$ and $\hat{{\bf n}}$.

\section{Theory}
We shall now demonstrate that the observed behavior can be well
understood using a generalization  of the approach that we have
previously  employed to quantitatively analyze the Dzyaloshinskii
transition in Ba$_{2}$CuGe$_2$O$_7$. While in our previous studies the
direction of the propagation vector remained constant,  which enabled us
to use an effectively 1-dimensional model, in the present experiment
${\bf q}$ rotates in the $(a,b)$ plane, and we have to  extend our
expression for the free energy of the spin structure to work in 2
dimensions.

\subsection{The continuous limit}
In general, any almost-antiferromagnetic spin structure can be described
in terms of the unit vector $\hat{{\bf m}}({\bf r})$ of local staggered
magnetization. This is the continuous-limit, when nearest- neighbor
spins are almost antiparallel and, in consequence, the magnetic
propagation vector is close to that of a N\'{e}el structure. For a
``relativistic'' spiral, such as the one in Ba$_{2}$CuGe$_2$O$_7$, the
continuous approximation is expected to work well when $|{\bf D}|$,
$\mu_{B} H
\ll J$, a condition well satisfied in our experiments.

The simplest form of the free energy that for $H=0$ gives a planar spin
spiral with propagation vector along $x$ and the spins rotating in the
$(x,z)$ plane, is $F
=
\rho_s/2 \int dxdy (\partial_x
\hat{{\bf m}}-\alpha/ \Lambda \ {\bf e}_y \times \hat{{\bf m}})^2$, where $\rho_s
\simeq J S^2$ is the temperature dependent spin  stiffness (see
Refs.~\onlinecite{ZM97-BACUGEO-L,ZM97-BACUGEO-B}), $\Lambda=a/\sqrt{2}$
($a$ being the lattice constant) is the nearest neighbor Cu-Cu distance,
and $\alpha$ is the angle by which spins are rotated at one step of the
spiral. For the ``relativistic'' spiral in Ba$_{2}$CuGe$_2$O$_7$,
$\alpha=-\arctan |{\bf D}|/J
\approx 10^{\circ}$. Of course, on a square lattice the spiral can
equally well propagate in the direction of the symmetrically equivalent
$y$-axis (with spins rotating in the $(y,z)$ plane), and a similar term
where $y$ is interchanged with $x$ has to be added to the free energy
functional. The purely ferromagnetic interactions between the Cu-layers
in Ba$_{2}$CuGe$_2$O$_7$ (Ref.~\onlinecite{Zheludev96-BACUGEO}) can be
accounted for by including the term $\rho_s\gamma
/2 \int dx dy \ (\partial_z \hat{{\bf m}})^2$, where $\gamma$ is the spin
stiffness anisotropy factor, $\gamma \approx 0.03$ for
Ba$_{2}$CuGe$_2$O$_7$. Finally, the interaction with external magnetic
field is given  by the Zeeman term $E_{Z}=-\int dx dy \ [
(\chi_{\bot}-\chi_{\|})({\bf H} \times \hat{{\bf m}})^2 /2  +
\chi_{\|} H^2/2 ]$ \ (for details see Refs.~\onlinecite{ZM97-BACUGEO-L,ZM97-BACUGEO-B}).
In this last expression  $\chi_{||}$ and $\chi_{\perp}$ are the
longitudinal and transverse local staggered susceptibilities of the
almost AF spin structure, respectively. For the classical spin model at
$T=0$ one gets $\chi_{\bot}=(g\mu_{B})^{2}/(8J\Lambda^{2})$ and
$\chi_{\|}=0$. Combining all of the above terms   gives us the simplest
expression for the free energy per Cu-plane, that is consistent  with
symmetric properties of the system and gives the ``right answer'' for
$H=0$:
\begin{eqnarray}
F= \int dxdy \ \left[ {\rho_s \over 2} \left((\partial_x \hat{{\bf m}}-
{\alpha
\over
\Lambda} {\bf e}_y \times \hat{{\bf m}})^2 \right. \right. \nonumber \\
\left. \left. + (\partial_y  \hat{{\bf m}}-{\alpha
\over \Lambda} {\bf e}_x
\times \hat{{\bf m}})^2 + \gamma (\partial_z \hat{{\bf m}})^2 \right)- \right. \nonumber \\
\left. -{(\chi_{\bot}-\chi_{\|})({\bf H} \times \hat{{\bf m}})^2 \over 2}  -
{\chi_{||} H^2 \over 2} \right] \label{free}
\end{eqnarray}

\subsection{Zero magnetic field.}
Let us first consider the system in the absence of a magnetic field.  We
want  to find a solution  for a {\it general} direction of the
propagation vector in the $(x,y)$ plane.  For the case when  ${\bf q}$
forms the angle $\phi$ with  the $x$-axis it is convenient to change the
coordinate system  to $(x',y',z')$ with $x'$ along the propagation
vector, $y'$ perpendicular to $x'$ in $(x,y)$ plane, and $z'= z$. The
derivatives and unit vectors are changed according to:
\begin{eqnarray}
\partial_x=\cos \phi \ \partial _{x'} - \sin \phi \ \partial _{y'} \\
\partial_y=\cos \phi \ \partial _{y'} + \sin \phi \ \partial _{x'}
\nonumber \\
{\bf e}_x=\cos \phi \ {\bf e}_{x'}- \sin \phi \ {\bf e}_{y'}\\ {\bf
e}_y=\cos \phi \ {\bf e}_{y'}+ \sin \phi \  {\bf e}_{x'} \nonumber
\end{eqnarray}
With these expressions Eq. (\ref{free}) after some algebra can be
rewritten as:
\begin{eqnarray}
F=\int dxdy \ \left[ {\rho_s \over 2} \left((\partial_{x'} \hat{{\bf m}})^2
+(\partial_{y'}  \hat{{\bf m}})^2 + \gamma (\partial_{z'} \hat{{\bf m}})^2 \right.
\right. -\\
-{2 \alpha \over \Lambda} ( \cos 2\phi \ {\bf e}_{y'} +
\sin 2\phi \ {\bf e}_{x'}) \cdot
\partial _{x'} \hat{{\bf m}} \times \hat{{\bf m}}- \nonumber\\
-{2 \alpha \over \Lambda} ( \cos 2\phi \ {\bf e}_{x'}+
\sin 2\phi \ {\bf e}_{y'}) \cdot
\partial _{y'} \hat{{\bf m}} \times \hat{{\bf m}} + \nonumber \\
\left. +{\alpha^2 \over \Lambda^2} \left.(1 + l_{z'}^2) \right)
-(\chi_{\bot}-\chi_{\|})({\bf H} \times \hat{{\bf m}})^2 /2  -
\chi_{\|} H^2/2 \right] \nonumber
\end{eqnarray}

For $H=0$, since we have selected $x'$  to be the direction of the
propagation vector, $\partial_{x'}\hat{{\bf m}} \ne 0$, while
$\partial_{y'}\hat{{\bf m}}=\partial_{z'}\hat{{\bf m}} \equiv 0$. It is
straightforward to verify that the above expression is minimized when i)
$\hat{{\bf m}}({\bf r})$ is periodic with the period $2
\pi
/\alpha$ and ii) along its vector of propagation $\hat{{\bf m}}$ uniformly
rotates in a plane that is perpendicular to $\cos 2\phi \ {\bf e}_{y'} +
\sin 2\phi \ {\bf e}_{x'}=
\cos \phi \ {\bf e}_{y} + \sin \phi \ {\bf e}_{x}$, i.e., the ``bisection
rule'' $\eta=\pi/2-\phi$ formulated in the previous section is
satisfied. In zero external field all spiral structures that conform
with this bisection rule are degenerate, i.e., they have the {\it same}
free energy.

\subsection{Non-zero in-plane field}
It is now easy to understand what happens in the case $H>0$.  From all
the possible spiral structures, energetically degenerate at $H=0$, the
system will pick the one that takes the most advantage of the Zeeman
energy $-({\bf H} \times \hat{{\bf m}})^{2}$, namely that which has its
spin plane normal to the field direction. Independent of the value of
$H$ (always assuming $\mu_{B} H\ll J$), instead of two equivalent
domains seen at $H=0$ one gets a single domain with $\phi=\pi/2-\psi$
and $\eta=\psi$. The particular case of ${\bf H} \|
 ({\bf e}_{x}+{\bf e}_{y})$ is of special interest. In this
case the propagation vector is also directed along $({\bf e}_{x}+{\bf
e}_{y})$ and one has a ``screw-type'' spiral with all spins
perpendicular to the propagation axis. Such a ``screw-type'' structure
is realized in MnSi and FeGe for arbitrary direction of ${\bf q}$.

While the above result accounts for the experimentally observed behavior
for $H\gtrsim 0.5$~T, we still have to explain why the rotations of the
spin plane and the propagation  vector are continuous and in small
fields some intermediate structure is realized. Obviously this is due to
some anisotropy effects that pick the propagation vector along $[1 1 0]$
(or $[1 \overline{1} 0]$) and the normal to the spin plane along $[1
\overline{1} 0]$ (or $[1 1 0]$) for $H=0$ in the first place. The most likely
source of anisotropy is spin-orbital interaction. For the present
discussion however, the actual origin of magnetic anisotropy is  of
little importance: we can take it into account by introducing a
phenomenological term into Eq.~(\ref{free}). This is done under the
assumption that the anisotropy energy $E_{A}$, as well as the Zeeman
energy $E_{Z}$, are much smaller than the energy scales of
Dzyaloshinskii or exchange interactions. Neither magnetic field, nor
anisotropy can distort any of the planar spiral structures in this
limit. Their only  effect is to pick the one that gives the greatest
gain in Zeeman and anisotropy energies.  $E_{A}$ can now be written as a
function of $\eta$ or $\phi$, which is in essence the same thing, since
for all the structures we are dealing with $\eta\equiv\pi/2-\phi$. Since
at $H=0$ we know from experiment that
$\eta=\pi/2$,\cite{Zheludev96-BACUGEO} the anisotropy term must be a
minimum at this point. It must also comply with the 4-fold symmetry of
the crystal, i.e., it must be invariant under $\eta
\rightarrow \eta +
\pi/2$. In the most general case it is written as $E_A=-A_{1}
\cos (4 \eta)-A_{2} \cos(8 \eta)\ldots$. It is also convenient to rewrite the
Zeeman energy in terms of $\phi$ and $\psi$:
$E_{Z}=-(\chi_{\bot}-\chi_{\|})({\bf H}
\times \hat{{\bf m}})^2 /2 = -(\chi_{\bot}-\chi_{\|})H^2 \cos^2 (\eta
-\psi)/2$. Minimizing $E_{Z}+E_{A}$ with respect to $\eta$ we obtain:
\begin{equation}
H^2= -{8A_1 \sin 4 \eta + 16A_2 \sin 8 \eta + \ldots \over
(\chi_{\bot}-\chi_{\|}) \sin 2(\eta-\psi)},\nonumber
\end{equation}
or, substituting $\eta=\pi/2-\phi$,
\begin{equation}
H^2= {8A_1 \sin 4 \phi + 16A_2 \sin 8 \phi + \ldots \over
(\chi_{\bot}-\chi_{\|}) \sin 2(\phi+\psi)}.\label{answer}
\end{equation}
This expression gives us the direction of the propagation vector and the
orientation of the spin rotation plane for arbitrary $H$.

\subsection{Comparison with experiment}
We  can now verify that the  analytical results obtained in the previous
section are  consistent with our experiments on Ba$_{2}$CuGe$_2$O$_7$.
To begin with, all the approximations that were made in the above
calculations are justified. Indeed, the anisotropy energy in
Ba$_{2}$CuGe$_2$O$_7$ corresponds to magnetic fields of several tenths
of a Tesla, fields in which the propagation vector actively rotates. The
energy of Dzyaloshinskii interactions, as previously mentioned,
corresponds to fields of the order of $2$~T, while the exchange energy
is roughly 10 times as large.

Equation~(\ref{answer}) can be directly used to fit the experimentally
measured $\phi(H)$. The solid lines in Fig.~\ref{angles} are the result
of a {\it global} fit of  this expression to all our data collected at
different $\psi$. The only adjustable parameter was the anisotropy
coefficient $\tilde{A}_{1}\equiv A_{1}/(\chi_{\bot}-\chi_{\|})$, while
we assumed $A_{2}=A_{3}=\ldots=0$. The refined value for $\tilde{A}_{1}$
is $1.95 (0.13) \cdot 10^{-3}$~T$^{2}$. The accuracy of the
single-parameter fit is quite remarkable.

It is interesting to note that our theory predicts qualitatively
different behaviors for the cases of $\psi=\pi/4$ and $\psi>\pi/4$,
which is particularly easy to understand in the case $A_2=A_3= \ldots
=0$. For $\psi=\pi/4$ Eq.~(\ref{answer}) turns into
$H^2=8A_1 \sin 4 \phi /(\chi_{\bot}-\chi_{\|})
 \sin 2 \eta = 16 A_1  \sin 2 \phi /(\chi_{\bot}-\chi_{\|})$.
We see that $\phi(H)$ changes continuously for $H^2 \leq 16
A_1/(\chi_{\bot}-\chi_{\|})$, has a kink at  this value and remains
constant (equal to $\pi/4$) above this threshold. On the other hand, for
$\psi > \pi/4$ the $\phi$ approaches $\pi/2-\psi$ asymptotically in high
fields, $\phi(H)$ is a smooth function and there is no threshold field.
Precisely this kind  of behavior is observed in experiment
(Fig.~\ref{angles}).

The limited data that we have for the orientation of the spin plane are
also totally consistent with theory. The only effect that our
theoretical model fails to account for is the observed slight variation
of $|{\bf q}|$. This phenomenon may be a result of corrections to the
continuous-limit approximation that we have ignored. Alternatively, it
may be a purely quantum effect, related to the small value ($S=1/2$) of
spins involved and the quasi-2-dimensional nature of the system.

\section{Concluding remarks}

Again we emphasize the key difference between FeGe and MnSi on  one
side, and Ba$_{2}$CuGe$_2$O$_7$ on the other. In the former two systems
the Dzyaloshinskii vector is pointing along the bond between interacting
spins, while in Ba$_{2}$CuGe$_2$O$_7$ ${\bf D}$ is orthogonal to this
bond.  As a result, in FeGe the normal to the spin plane $\hat{{\bf n}}$, and
the propagation vector ${\bf q}$ are always collinear. In contrast, in
Ba$_{2}$CuGe$_2$O$_7$ this colinearity is replaced by the ``bisection
rule'': ${\bf q}$ and $\hat{{\bf n}}$ form equal angles with the $a$ axis of
the crystal, but are not colinear.

In conclusion, we believe that the present  work  gives a fairly
complete picture of the rather exotic {\it static} magnetic properties
of the square-lattice Dzyaloshinskii-Moriya antiferromagnet
Ba$_{2}$CuGe$_2$O$_7$. Much remains to be learned in the study of
magnetic critical behavior, as well as the dynamical properties of the
soliton lattice, realized in finite fields applied along the $c$-axis.

\acknowledgements
This study was supported in part by NEDO
   (New Energy and Industrial Technology Development Organization)
   International Joint Research Grant and the U.S. -Japan Cooperative
   Program on Neutron Scattering.
We thank Scott Moore and Brent Taylor for expert technical assistance.
Oak Ridge National Laboratory is managed for the U.S. D.O.E. by Lockheed
Martin Energy Research Corporation under contract DE-AC05-96OR22464.
Work at Brookhaven National Laboratory  was carried out under Contract
No. DE-AC02-76CH00016, Division of Material Science, U.S. D.O.E.

%\bibliographystyle{prsty}
%\bibliography{/home/zhelud/bib/spiral}

\begin{figure}
\caption{Contour plots of elastic neutron scattering intensity measured
in Ba$_{2}$CuGe$_2$O$_7$ around the $(1,0,0)$ antiferromagnetic
zone-center for three values of magnetic field applied along the $[1 0
0]$ direction. Plot (a) was measured for a zero-field-cooled sample, and
field-cooling was used for plots (b) and (c).}
\label{cont}
\end{figure}

\begin{figure}
\caption{Experimental geometry: The magnetic field ${\bf H}$ is applied
in the $(a,b)$ crystallographic plane at an angle $\psi$ to the $[1 1
0]$ direction. The propagation vector ${\bf q}$ and the normal to the
plane of spin rotations form angles $\phi$ and $\eta$ with the $[1 1 0]$
direction, respectively.}
\label{notation}
\end{figure}

\begin{figure}
\caption{Field-dependence of the direction of magnetic propagation vector
measured in Ba$_{2}$CuGe$_2$O$_7$ for several directions of magnetic
field applied in the $(a,b)$ crystallographic plane. The solid lines
represent a theoretical fit to the data, described in the text.}
\label{angles}
\end{figure}

\begin{figure}
\caption{$\phi$, the direction of magnetic propagation vector,
measured in Ba$_{2}$CuGe$_2$O$_7$ as a function of $\psi$, the direction
of the external magnetic field for $H=2$~T. The  dashed line shows the
theoretical result.}
\label{linear}
\end{figure}

\begin{figure}
\caption{Measured trajectories traced by the magnetic propagation vector in
Ba$_{2}$CuGe$_2$O$_7$ as a magnetic field is applied in the $(a,b)$
plane of the crystal. The solid lines are guides for the eye.}
\label{trajectory}
\end{figure}

\begin{figure}
\caption{Measured intensities of magnetic satellites around the $(1,0,0)$ and
$(0,1,0)$ antiferromagnetic zone-centers in Ba$_{2}$CuGe$_2$O$_7$
plotted as a function of magnetic field applied along the $a$-axis. The
solid line are the theoretically  predicted dependences, as described in
the text.}
\label{intensity}
\end{figure}

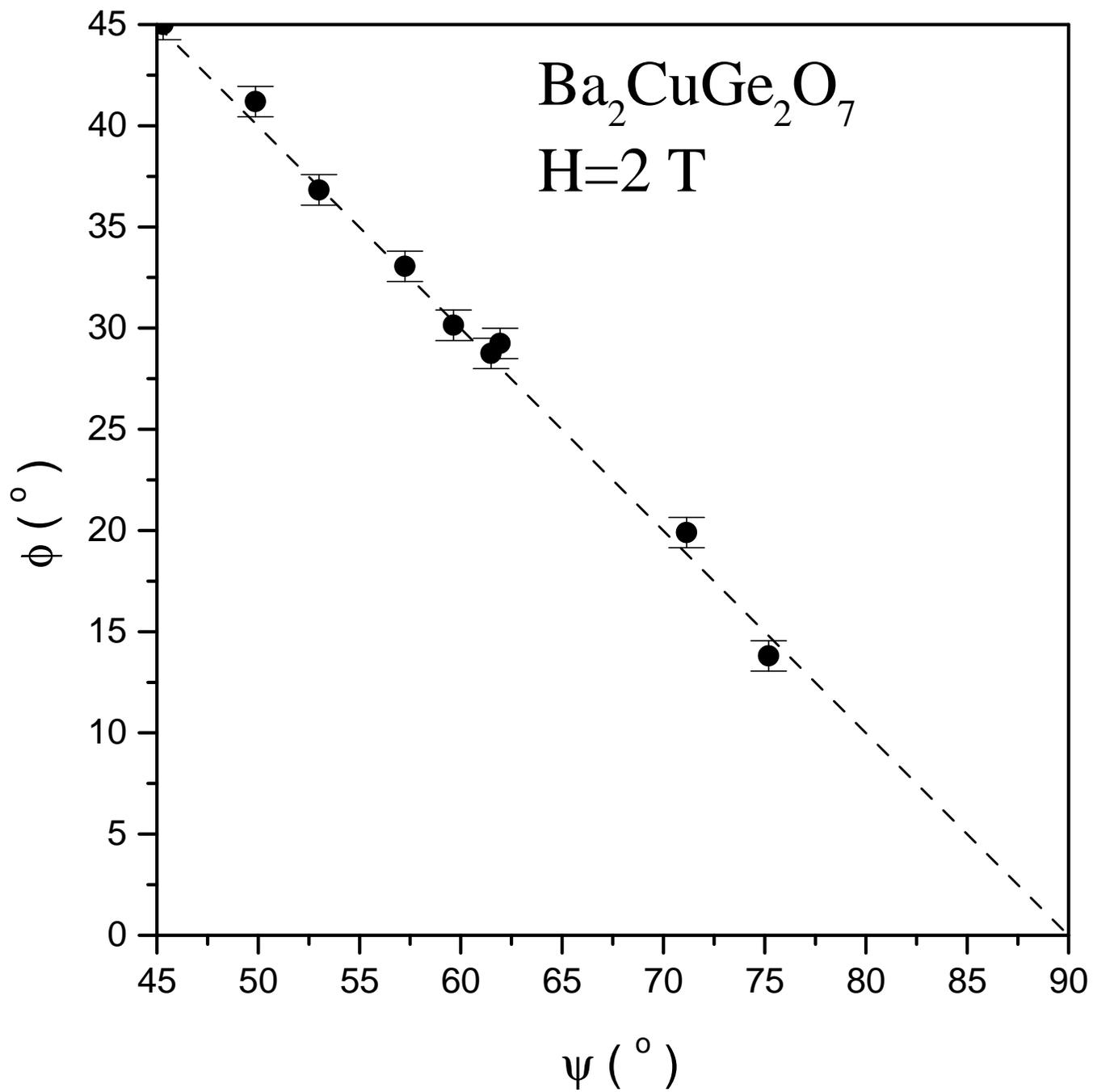
\end{document}